\documentclass{article}
\usepackage{graphicx}
\usepackage{url}
\usepackage{amsmath}
\usepackage[english]{babel}
\usepackage{authblk}
\usepackage{hyperref}
\usepackage{amssymb}

\title{\textbf{Redactable Blockchains: An Overview}}

\author[1]{Federico Calandra}
\author[1]{Marco Bernardo}
\author[1]{Andrea Esposito}
\author[2]{Francesco Fabris}

\affil[1]{Dipartimento di Scienze Pure e Applicate, University of Urbino, Italy}
\affil[2]{Dipartimento di Matematica, Informatica e Geoscienze, University of Trieste, Italy}

\date{}

\begin{document}

\maketitle

%%%%%%%%%%%%%%%%%%%%%%%%%%%%%%%%%%%%%%%%%%%%%%%%%%%%%%%%%%%%%%%%%
%                                                               %
%                                                               %
% Abstract                                                      %
%                                                               %
%                                                               %
%%%%%%%%%%%%%%%%%%%%%%%%%%%%%%%%%%%%%%%%%%%%%%%%%%%%%%%%%%%%%%%%%

\begin{abstract}
\noindent
Blockchains are widely recognized for their immutability, which provides robust guarantees of data integrity
and transparency. However, this same feature poses significant challenges in real-world situations that
require regulatory compliance, correction of erroneous data, or removal of sensitive information. Redactable
blockchains address the limitations of traditional ones by enabling controlled, auditable modifications to
blockchain data, primarily through cryptographic mechanisms such as chameleon hash functions and alternative
redaction schemes. This report examines the motivations for introducing redactability, surveys the
cryptographic primitives that enable secure edits, and analyzes competing approaches and their shortcomings.
Special attention is paid to the practical deployment of redactable blockchains in private settings, with
discussions of use cases in healthcare, finance, Internet of drones, and federated learning. Finally, the
report outlines further challenges, also in connection with reversible computing, and the future potential
of redactable blockchains in building law-compliant, trustworthy, and scalable digital infrastructures.
\end{abstract}

%%%%%%%%%%%%%%%%%%%%%%%%%%%%%%%%%%%%%%%%%%%%%%%%%%%%%%%%%%%%%%%%%
%
%
\section{Introduction}
\label{Sec:Intro}
%
%
%%%%%%%%%%%%%%%%%%%%%%%%%%%%%%%%%%%%%%%%%%%%%%%%%%%%%%%%%%%%%%%%%

Blockchain technology is a form of decentralized digital ledger that securely records transactions across a
network of computers. First introduced with Bitcoin~\cite{Bitcoin_A_Peer_to_Peer} in~2008, blockchains allow
participants to reach consensus and trust data without relying on central authorities. Pieces of data,
called transactions, are grouped into blocks, each of which is linked to the previous one so as to form a
chain that is transparent, secure, and resistant to tampering. Today, blockchains power not only
cryptocurrencies but are also used in finance, supply chain management, healthcare, digital identity, and
many other fields where trust and verifiability are crucial.

At the heart of blockchain technology lies a core property: \emph{immutability}. Once information is added
to a blockchain, it is designed to stay there forever. No single participant can alter or delete data on
their own. This feature ensures that historical records are protected from manipulation and builds user
trust in the system's integrity. However, this strength can become a weakness in real-world situations.

This challenge is known as the \emph{immutability dilemma}~\cite{Resolving_Blockchains_Immutable_Dilemma}.
As blockchains become used for more applications, completely unchangeable data can create serious problems.
For example, current laws such as the European Union’s General Data Protection Regulation (GDPR) give
individuals the ``right to be forgotten''~\cite{Art17_GDPR}, i.e., the ability to have their personal data
deleted on request. Traditional blockchains, by design, cannot meet such requirements. Moreover, if
incorrect, illegal, or harmful information is added to a blockchain, there is no way to remove it. This not
only raises legal and ethical concerns, but also threatens to limit blockchain adoption in sensitive areas
such as healthcare, finance, and public services. Additionally, under the European Union’s Digital Services
Act (DSA)~\cite{EU_DSA2022}, platforms that fail to provide mechanisms for the timely removal of illegal or
harmful content may face sanctions or even be suspended from operating within the EU. This further
emphasizes the need for data modification capabilities in blockchain architectures.

To address the aforementioned issues, researchers have proposed the introduction of \emph{redactability}
into blockchain systems~\cite{Redactable_Blockchain_or_Rewriting}. Redactable blockchains are designed to
allow specific, controlled edits or deletions of data -- but only under strict rules and with strong
security guarantees. Thanks to advanced cryptographic tools, such as chameleon hash
functions~\cite{Chameleon_hashes_with_ephemeral}, it is now possible to make limited changes to blockchain
data while preserving the chain’s structure, transparency, and tamper-evident qualities.

This paper explores the technical foundations of state-of-the-art approaches that make redactable
blockchains possible. After discussing why redactability is important
(Section~\ref{Sec:Immutability_Dilemma}), we present the cryptographic techniques that enable it in public
blockchains like chamaleon hashing (Section~\ref{Sec:Chamaleon_Hash}) and others
(Section~\ref{Sec:Alternative_Techniques}), along with their major limitations
(Section~\ref{Sec:Limitations}). We then see how redactability can be incorporated into private blockchains
(Section~\ref{Sec:Private_Blockchains}) and examine several concrete applications and use cases where those
techniques can be impactful (Section~\ref{Sec:Applications}). We finally conclude the report by summarizing
the main insights and outlining directions for future work (Section~\ref{Sec:Conclusion}).

%%%%%%%%%%%%%%%%%%%%%%%%%%%%%%%%%%%%%%%%%%%%%%%%%%%%%%%%%%%%%%%%%
%
%
\section{The Immutability Dilemma}
\label{Sec:Immutability_Dilemma}
%
%
%%%%%%%%%%%%%%%%%%%%%%%%%%%%%%%%%%%%%%%%%%%%%%%%%%%%%%%%%%%%%%%%%

Blockchain technology is fundamentally designed with immutability as a core attribute, meaning that once
data or transactions are recorded on the ledger, they are virtually impossible to alter or remove. This
feature ensures the integrity, transparency, and trustworthiness of the data. However, this very strength
has given rise to the \emph{immutability dilemma}, presenting significant challenges that hinder the broader
applicability and development of blockchain
systems~\cite{Redactable_blockchains_with_integer,Redactable_Blockchain_Based_on,Redactable_Blockchain_or_Rewriting,Redactable_Blockchain_in_Decentralized}.

In particular, immutability becomes problematic in real-world scenarios that demand flexibility,
accountability, and adaptability. Several critical concerns have emerged, which demonstrate the limitations
of an unchangeable ledger in modern digital ecosystems:

\pagebreak

\begin{itemize}

  \item \textbf{Privacy and Regulatory Compliance (``Right to be Forgotten''):} A primary concern is the
conflict with evolving legislative frameworks, most notably the European Union’s
GDPR~\cite{Resolving_Blockchains_Immutable_Dilemma,Redactable_Blockchain_Based_on,A_Survey_on_Redactable,PRBFPT_A_Practical_Redactable,Redactable_blockchain_with_anonymity}.
The GDPR establishes the ``right to be forgotten'' (RTBF), which grants individuals the authority to request
the deletion of their personal information. Immutable blockchains cannot inherently comply with these
mandates, potentially leading to substantial fines for companies and forcing honest users to withdraw from
systems if no recourse exists for data removal. Even encrypting data may not solve this issue, as keys can
be leaked, and metadata alone can often reveal private details~\cite{Redactable_Blockchain_or_Rewriting}.
  
  \item \textbf{Management of Erroneous or Malicious Content:} The append-only nature means that erroneous,
malicious, or inappropriate content, once recorded, becomes irreversible and permanently stored. Examples
include copyright-infringing content, sensitive personal information, pornographic materials, or even
malicious links and malware~\cite{A_quantitative_analysis_of}. This poses a challenge for law enforcement
and negatively impacts the blockchain ecosystem, as users may be unwilling to broadcast or store such an
illicit content~\cite{A_quantitative_analysis_of,Child_porn_on_bitcoin}. Moreover, design flaws in the
blockchain protocol itself can lead to the unintended inclusion of incorrect or undesirable data. In these
cases, the inability to modify the ledger prevents corrective actions and may result in systemic
vulnerabilities or loss of trust in the platform.
  
  \item \textbf{Storage Overhead and Scalability Issues:} Maintaining an ever-expanding log of all
transactions in high-load systems (potentially thousands of records per second) leads to a serious problem
of storing and processing vast volumes of information. The perpetual storage of data on an unprunable
blockchain, as well as the exponential growth of participants, raise significant storage concerns and can
affect performance. For instance, the Bitcoin ledger size reached 650~GB in July
2025~\cite{Size_of_the_bitcoin_blockchain}.
  
  \item \textbf{Vulnerabilities in Smart Contracts:} The immutability of smart contracts means that even if
vulnerabilities or flaws are discovered in their code, they cannot be fixed once deployed, leading to
ongoing security issues. A notable example is the DAO attack in 2016, which resulted in the loss of
approximately 40 million USD in Ether and caused a contentious hard fork, splitting the Ethereum
community~\cite{Ethereum_A_Next_Generation, Smart_contract_development_challenges}. Amending or patching
contract code by merely appending new versions is inefficient and wastes resources.

\end{itemize}

Motivated by these profound challenges, the concept of redactable blockchain has emerged as a novel
solution. The core idea is to introduce a controlled degree of data modifiability into blockchain systems
while simultaneously preserving their fundamental principles of security, transparency, and
decentralization.

A primary motivation is the need to ensure compliance with data protection regulations and provide
mechanisms for removing malicious or incorrect data, both of which are difficult to achieve in traditional
immutable systems. Redactable blockchains allow authorized modifications that preserve ledger integrity
while enabling legal and operational flexibility.

Additionally, applications in sectors like healthcare, finance, and IoT often require the ability to update
or correct recorded data. Redactability addresses this need by supporting a more adaptable and efficient
storage model.

In essence, the development of redactable blockchain technologies represents a pragmatic response to
real-world demands, bridging the gap between theoretical immutability and practical functionality. This
evolution expands the applicability of blockchain systems beyond static environments, making them more
viable in complex, regulated, and data-intensive contexts.

To achieve these goals, redactable blockchains often leverage advanced cryptographic techniques. The most
common approach involves replacing traditional cryptographic hash functions with chameleon hash
functions~\cite{Chameleon_hashes_with_ephemeral}, which allow authorized entities with a secret trapdoor key
to find collisions and modify block content without altering the block's hash value or breaking the
integrity of the chain. Other methods include using zero-knowledge
proofs~\cite{Non_interactive_zero_knowledge} for verifiable data erasure and novel structures based on
integer-valued polynomials~\cite{Redactable_blockchains_with_integer} or verifiable delay
functions~\cite{Redactable_Blockchain_Based_on}.

%%%%%%%%%%%%%%%%%%%%%%%%%%%%%%%%%%%%%%%%%%%%%%%%%%%%%%%%%%%%%%%%%
%
%
\section{Chamaleon Hashing in Public Blockchains}
\label{Sec:Chamaleon_Hash}
%
%
%%%%%%%%%%%%%%%%%%%%%%%%%%%%%%%%%%%%%%%%%%%%%%%%%%%%%%%%%%%%%%%%%

Chameleon hashing (CH) is a core cryptographic technique that plays a pivotal role in supporting redactable
blockchains, addressing the inherent immutability of traditional blockchain systems. CH enables data
modification on a blockchain without breaking the cryptographic links between consecutive blocks. Instead of
recalculating the entire proof of work, CH allows a hash collision to be found for new data so as to match
the original hash.

A CH function is a special collision-resistant hash function that includes a trapdoor. This unique property
allows hash collisions to be efficiently generated when the secret trapdoor key is known. Without the
trapdoor, finding collisions remains computationally infeasible, similar to a standard hash function.

A standard CH scheme typically involves four sub-algorithms:

\begin{enumerate}

  \item \textbf{Key Generation:} Takes a security parameter as input and outputs a public key and a secret
trapdoor key.

  \item \textbf{Hashing:} Takes the public key, a message, and a random number (or implicitly generates
randomness) to produce a hash value.

  \item \textbf{Verification:} Inputs the public key, a message, a random number, and a hash value, then
returns a boolean indicating validity.

  \item \textbf{Collision Finding/Adaptation:} Given the trapdoor key, an old message, its random number,
its hash value, and a new message, it returns a new random number that produces the \textit{same} hash value
for the new message.

\end{enumerate}

The rest of this section explores the structure and application of CH functions.
Section~\ref{Subsec:Foundation_of_CH} presents their formal definition, properties, and cryptographic
foundations, including a general construction based on claw-free trapdoor permutations.
Section~\ref{Subsec:Blockchain_redaction_CH} explains how these functions are integrated into blockchain
architectures to enable efficient and secure redactions. Finally, Section~\ref{Subsec:Security_of_CH}
discusses the various levels of security guarantees that CH functions can provide and highlights which
variants are most suitable for redactable blockchains.

%%%%%%%%%%%%%%%%%%%%%%%%%%%%%%%%%%%%%%%%%%%%%%%%%%%%%%%%%%%%%%%%%
%
\subsection{Chameleon Hash Functions}
\label{Subsec:Foundation_of_CH}
%
%%%%%%%%%%%%%%%%%%%%%%%%%%%%%%%%%%%%%%%%%%%%%%%%%%%%%%%%%%%%%%%%%

CH functions form the cryptographic backbone of redactable blockchain architectures by enabling collision
generation under controlled conditions. Formally introduced by Krawczyk and
Rabin~\cite{Chameleon_Signatures}, a CH function is a collision-resistant hash function for anyone except a
party holding a special piece of trapdoor information. This subsection explores the mathematical
construction of CH functions.

To understand how CH functions enable controlled redactions in blockchains, we begin by formalizing their
fundamental properties and security guarantees (Section~\ref{Subsec:General_Prop_of_CH}). We then describe a
general construction based on claw-free trapdoor permutations, which serves as a blueprint for building
secure and efficient CH functions (Section~\ref{Subsec:Construction_CH}). Finally, we explore how these
functions can be optimized for practical use by composing them with standard hash functions to enhance
efficiency and composability (Section~\ref{Subsec:Composability_of_CH}).

%%%%%%%%%%%%%%%%%%%%%%%%%%%%%%%%%%%%%%%%%%%%%%%%%%%%%%%%%%%%%%%%%
\subsubsection{Fundamental Properties}
\label{Subsec:General_Prop_of_CH}
%%%%%%%%%%%%%%%%%%%%%%%%%%%%%%%%%%%%%%%%%%%%%%%%%%%%%%%%%%%%%%%%%

Let $R$ be a recipient who publishes a public hashing key $\mathsf{HK}_R$ and holds a corresponding secret
trapdoor key $\mathsf{CK}_R$. A CH function $\mathsf{cham\text{-}hash}_R(m, r)$ maps a message $m$ and a
random value $r$ to a hash output $h$ in a way that the following properties are satisfied:

\begin{itemize}

  \item \textbf{Collision Resistance:} Without knowledge of $\mathsf{CK}_R$, it is computationally
infeasible to find $(m_1, r_1)$ and $(m_2, r_2)$ such that:
$$m_1 \ne m_2 \textrm{ and } \mathsf{cham\text{-}hash}_R(m_1, r_1) = \mathsf{cham\text{-}hash}_R(m_2, r_2)$$

  \item \textbf{Trapdoor Collisions:} With knowledge of $\mathsf{CK}_R$, for any $m_1, r_1$, and $m_2$ one
can efficiently compute $r_2$ such that:
$$\mathsf{cham\text{-}hash}_R(m_1, r_1) = \mathsf{cham\text{-}hash}_R(m_2, r_2)$$

  \item \textbf{Uniform Output Distribution:} The distribution of outputs for randomly chosen $r$ is
independent of the message $m$, thus preventing information leakage from the hash value.

\end{itemize}

%%%%%%%%%%%%%%%%%%%%%%%%%%%%%%%%%%%%%%%%%%%%%%%%%%%%%%%%%%%%%%%%%
\subsubsection{General Construction via Claw-Free Trapdoor Permutations}
\label{Subsec:Construction_CH}
%%%%%%%%%%%%%%%%%%%%%%%%%%%%%%%%%%%%%%%%%%%%%%%%%%%%%%%%%%%%%%%%%

A general and elegant construction of CH functions can be derived from claw-free trapdoor permutations. Let
$(f_0, f_1)$ be a pair of permutations over a common finite\footnote{The domain $\mathcal{D}$ is necessarily
finite in practical constructions. For instance, in RSA-based implementations, $\mathcal{D} =
\mathbb{Z}_n^*$ where $n$ is the RSA modulus.} domain $\mathcal{D}$. A pair of permutations $(f_0, f_1)$ is
called \emph{claw-free} if it is computationally infeasible to find values $x$ and $y$ in $\mathcal{D}$ such
that $f_0(x) = f_1(y)$. Each $f_i$ must be invertible given some trapdoor information.

Let $m = m[1] \ldots m[k] \in \{ 0, 1 \}^k$ be a binary message whose representation is suffix-free, i.e.,
the representation never yields two messages such that one is a suffix of the other\footnote{Suffix freeness
ensures unambiguous parsing of the message bits, thus preventing different interpretations that would lead
to different hash computations.}. Then, for a random seed $r \in \mathcal{D}$, the CH function is computed
by applying the permutations $f_{m[i]}$ for $1 \le i \le k$, where $f_{m[i]} = f_0$ \linebreak if $m[i] = 0$
and $f_{m[i]} = f_1$ if $m[i] = 1$, in the following order:

\begin{equation}
  \mathsf{cham\text{-}hash}(m, r) = f_{m[k]} \circ f_{m[k-1]} \circ \ldots \circ f_{m[1]}(r)
\end{equation}

\noindent
which ensures the aforementioned cryptographic properties:

\begin{itemize}

  \item \textbf{Collision Resistance:} If the trapdoor is unknown, the function remains collision-resistant
due to the claw-free property of $(f_0, f_1)$. Specifically, if an adversary could find $m_1 \ne m_2$ and
$r_1, r_2$ such that:

\begin{equation*}
  \mathsf{cham\text{-}hash}(m_1, r_1) = \mathsf{cham\text{-}hash}(m_2, r_2)
\end{equation*}

\noindent
then this would imply the ability to find a claw in $(f_0, f_1)$, contradicting the claw-free assumption.

  \item \textbf{Trapdoor Collisions:} A holder of the trapdoor (i.e., the ability to invert both $f_0$ and
$f_1$) can efficiently find a collision. Given $(m_1, r_1)$ and any desired message $m_2$, one can compute
$r_2$ such that:

\begin{equation*}
  \mathsf{cham\text{-}hash}(m_1, r_1) = \mathsf{cham\text{-}hash}(m_2, r_2)
\end{equation*}

\noindent
by sequentially applying the inverse functions in reverse order over the new message $m_2$.

  \item \textbf{Uniform Output Distribution:}
The output distribution of the CH function is uniform over the codomain for uniformly chosen $r \in
\mathcal{D}$ because each $f_{m[i]}$ is a permutation. This uniformity ensures that no information about the
input message $m$ is leaked through the hash value alone.

\end{itemize}

This general construction can be efficiently instantiated by using concrete number-theoretic assumptions.
One such realization is based on the hardness of integer factorization, where the permutations $f_0$ and
$f_1$ are defined by using modular squaring and multiplication in an RSA-like setting. Another well-known
instantiation employs the discrete logarithm problem, where the permutations operate as exponentiations in a
cyclic group of prime order. Both constructions allow efficient trapdoor inversion while ensuring the
claw-free property necessary for CH.

%%%%%%%%%%%%%%%%%%%%%%%%%%%%%%%%%%%%%%%%%%%%%%%%%%%%%%%%%%%%%%%%%
\subsubsection{Composability and Efficiency}
\label{Subsec:Composability_of_CH}
%%%%%%%%%%%%%%%%%%%%%%%%%%%%%%%%%%%%%%%%%%%%%%%%%%%%%%%%%%%%%%%%%

Due to performance considerations, it is often desirable to first hash an arbitrary-length message $m$ by
using a fast collision-resistant hash function (e.g., SHA-256), thus producing a short digest $h_m$. Then
the CH function is applied to $h_m$ instead of the full message:

\begin{equation}
  \mathsf{cham\text{-}hash}(h_m, r)
\end{equation}

\noindent
This layered approach preserves collision-resistance and enables practical efficiency, making CH functions
highly suitable for real-world redactable blockchain implementations.

%%%%%%%%%%%%%%%%%%%%%%%%%%%%%%%%%%%%%%%%%%%%%%%%%%%%%%%%%%%%%%%%%
%
\subsection{CH Blocks Redactability}\label{Subsec:Blockchain_redaction_CH}
%
%%%%%%%%%%%%%%%%%%%%%%%%%%%%%%%%%%%%%%%%%%%%%%%%%%%%%%%%%%%%%%%%%

Traditional blockchains enforce immutability through cryptographic hashing, where each block's hash depends
on its content and the hash of the previous block. Even a minor change to transaction data within a block
alters its hash, causing a cascading effect that invalidates all subsequent blocks in the chain, thereby
breaking its integrity. This makes post-recording modification virtually impossible without a hard fork.

Redactable blockchains overcome this limitation by integrating CH functions into their structure, primarily
by replacing the conventional hash function used for linking blocks or for constructing Merkle trees of
transactions within blocks. Here is a detailed breakdown of the redaction process using CH:

\begin{enumerate}

  \item \textbf{Modification of Block Structure:} The blockchain's block header is typically extended to
include fields for CH randomness (or check value) and potentially the public key. For example, the inner
hash function used to summarize block data before mining, or the transaction hash function within a Merkle
tree, is replaced by a CH function.

  \item \textbf{Transaction/Block Hashing with CH:} When a new transaction or block is created, its hash
value $h$ is generated alongside a corresponding random number $r$. The value $h$ is then typically
incorporated into the Merkle tree root of the block or directly into the block header's hash linkage.

  \item \textbf{The Redaction Event:} Suppose that a specific transaction or block content TX needs to be
modified -- e.g., due to legal requirements or to correct erroneous data.

  \item \textbf{Collision Computation (The Core of Redaction):} The entity holding the trapdoor key for the
CH function used for that specific data initiates the redaction. They input the original message TX, its
associated randomness $r$, the target hash $h$, and the new modified message TX$'$ into the collision finder
algorithm. This algorithm generates a new randomness $r'$ such that hashing TX$'$ with $r'$ yields the same
hash $h$.

  \item \textbf{Block/Transaction Update:} The original transaction $(\text{TX}, r)$ in the block is
replaced with the modified version $(\text{TX}', r')$. Crucially, because the hash value $h$ remains
unchanged, the Merkle root and block linkage are preserved, avoiding cascading hash recalculations.

  \item \textbf{Broadcast and Consensus:} The modified block is broadcast to the network. Participating
nodes verify the new block. Since the hash value remains consistent, the chain's integrity is maintained and
nodes can adopt the updated chain in accordance with pre-agreed redaction rules -- even if it replaces a
longer version of the chain.

\end{enumerate}

This mechanism enables controlled data manipulation (modification or deletion) and offers improved
computational performance and storage efficiency by eliminating the need for cascading hash recalculations.

%%%%%%%%%%%%%%%%%%%%%%%%%%%%%%%%%%%%%%%%%%%%%%%%%%%%%%%%%%%%%%%%%
%
\subsection{CH Security Properties}
\label{Subsec:Security_of_CH}
%
%%%%%%%%%%%%%%%%%%%%%%%%%%%%%%%%%%%%%%%%%%%%%%%%%%%%%%%%%%%%%%%%%

Different levels of collision resistance define the security guarantees of CH functions:

\begin{enumerate}

  \item \textbf{Weak Collision Resistance (w-CR):} This is the most basic form of resistance, ensuring that
finding collisions is hard without the trapdoor~\cite{Chameleon_Signatures}. However, many w-CR schemes
suffer from the \emph{key-exposure problem}, where observing even a single collision may compromise the
trapdoor, allowing unauthorized modifications. This renders them unsuitable for redactable blockchains,
where collisions are intentionally created. To mitigate this, \emph{key-exposure free} CH schemes have
been developed, which prevent trapdoor recovery even after multiple collisions are revealed.

  \item \textbf{Enhanced Collision Resistance (e-CR):} This strengthens w-CR by ensuring that an adversary
cannot find a collision for a specific hash value, provided no collision for that hash has been publicly
exposed~\cite{Redactable_Blockchain_or_Rewriting}. While stronger than w-CR, some studies suggest that it
may still be insufficient in certain redactable blockchain scenarios.

  \item \textbf{Standard Collision Resistance (s-CR):} It guarantees that finding a collision is infeasible
if no collision involving the target message has ever been
revealed~\cite{Chameleon_hashes_with_ephemeral}. Similar to e-CR, its adequacy for all redactable blockchain
applications is still under discussion.

  \item \textbf{Full Collision Resistance (f-CR):} This is the strongest known notion for CH security. It
combines the properties of both e-CR and s-CR, ensuring that finding a collision for a specific
\textit{hash-message pair} is infeasible unless a collision for that exact pair has already been
exposed~\cite{Bringing_order_to_chaos}.

\end{enumerate}

All these levels also aim for \emph{indistinguishability}, which ensures that the randomness generated by
the initial hashing algorithm is cryptographically indistinguishable from the randomness generated by the
adaptation algorithm. \linebreak In this way, an adversary is prevented from determining whether a
transaction has been modified by simply inspecting its randomness.

%%%%%%%%%%%%%%%%%%%%%%%%%%%%%%%%%%%%%%%%%%%%%%%%%%%%%%%%%%%%%%%%%
%
%
\section{Alternative Redactability Techniques}
\label{Sec:Alternative_Techniques}
%
%
%%%%%%%%%%%%%%%%%%%%%%%%%%%%%%%%%%%%%%%%%%%%%%%%%%%%%%%%%%%%%%%%%

Beside the use of CH functions, several other techniques have been proposed to support redactable
blockchains.

\paragraph{Polynomial-Based Redaction.}
This technique uses polynomials to structure and link blockchain data, thus enabling modification without
relying on hash recalculations~\cite{Redactable_blockchains_with_integer}. A notable variant employs
integer-valued polynomials, where data changes are integrated by altering coordinates and adjusting a
padding field. This method offers fine-grained control over modifications and supports dynamic contexts such
as finance and healthcare, thanks to efficient integer-based operations and tunable security. However,
reliance on finite fields can limit scalability and the method is often unsuitable for proof-of-work
blockchains. In addition, it may not preserve previous block states and usually requires substantial changes
to the blockchain structure, hindering compatibility with mainstream systems~\cite{A_Survey_on_Redactable}.

\paragraph{RSA-Based Redaction.}
%ZZZ (di quelle private non abbiamo ancora parlato)
%Primarily applied to private blockchains,
This method builds immutability on the computational hardness of
the RSA problem~\cite{RSA_and_redactable_blockchains}. Blocks consist of a permanent prefix, content, and a
redactable suffix, with content linked forward using a one-way function. It is computationally efficient for
some operations and offers corruption resistance, but typically depends on a central authority to approve
redactions. This raises concerns about decentralization and auditability, especially because modification
history may not be preserved. Furthermore, this approach can be vulnerable to attacks that exploit redaction
privileges and is inefficient for real-time contexts.

\paragraph{Data-Appending-Based Redaction.}
Instead of altering existing data, this approach appends updates as new transactions or data elements,
maintaining the entire historical record~\cite{Mchain_how_to_forget}. Security is ensured through consensus
mechanisms that validate these appended changes. While this method supports robust integrity verification
over time, it does not truly remove data. Original content, even if legally problematic, remains on-chain
and accessible. The resulting accumulation of data can negatively impact scalability and storage efficiency.
Moreover, managing keys for encrypted historical data adds further complexity.

\paragraph{Voting-Based (Consensus-Based) Redaction.}
In this model, redactions are approved via a consensus process involving miners or committee members who
evaluate requests based on pre-established policies~\cite{Redactable_blockchain_permissionless}. This can be
implemented using extended block structures or parallel ``Redaction'' and ``Standard'' chains. It introduces
multi-party control and can enhance accountability by preserving the hash of original data, allowing public
verification. Nonetheless, the voting process often incurs delays and may require protocol or structural
changes that are incompatible with popular blockchains like Bitcoin or Ethereum. Its security depends
heavily on the honesty of participating entities and may increase the system's computational and bandwidth
demands.

\paragraph{Local Redaction (Functionality-Preserving Local Erasures).}
This approach modifies only the local storage of individual nodes, removing or garbling sensitive
information without altering the global chain state~\cite{Erasing_data_blockchain_nodes}. Nodes are allowed
to store different local views, provided that they agree on the chain’s logical history. Although this
satisfies some regulatory demands, it undermines decentralization by creating inconsistencies among nodes
and weakens verification by relying on heuristic assumptions. Additionally, since the original data may
remain on other nodes, the effectiveness of redaction is limited.

\paragraph{Hard Forks.}
A hard fork constitutes a deliberate protocol change that splits the blockchain, creating a new version that
omits or modifies prior data. This was famously used in Ethereum to address the DAO incident. It offers a
direct solution for historical data alteration but at the cost of major disruption. Forks require extensive
coordination, often compromise decentralization, and are generally unfeasible for large, established
blockchains.

\paragraph{Zero-Knowledge Proofs NIZKs and zk-SNARKs.}
These cryptographic techniques enable the validation of a claim without revealing its
content~\cite{Non_interactive_zero_knowledge}. In redactable blockchains, they can prove that a redaction
complies with predefined policies without exposing original
data~\cite{Decentralised_Redactable_Blockchain_A}. They support anonymity, prevent key leakage, and allow
for selective removal of non-executable transaction components, mitigating the cascade of dependent changes.
Despite these benefits, such schemes are computationally intensive, complex to implement, and must be
carefully designed to ensure soundness and resistance to malleability attacks\footnote{Malleability attacks
exploit the ability to modify a cryptographic proof while maintaining its apparent validity, potentially
allowing unauthorized reuse or manipulation of redaction authorizations. Such attacks can compromise the
integrity of redaction policies by enabling attackers to derive valid proofs for unauthorized modifications
from legitimate redaction proofs.}.

\pagebreak

\paragraph{Verifiable Delay Functions (VDFs).}
VDFs are cryptographic primitives that require a prescribed amount of time to compute and produce a proof
that this time has elapsed~\cite{Verifiable_delay_functions}. VDF-based solutions attach a time-elapse proof
to each block, ensuring that a certain amount of time has passed~\cite{Redactable_Blockchain_Based_on}.
Redaction is achieved through the rapid construction of chain forks, with the VDF ensuring enforced
synchronization of these redactions to maintain blockchain consistency. However, current implementations
often rely on centralized control of a trapdoor, which compromises decentralization. They also tend to
impose high communication overheads, limiting practical deployment.

%%%%%%%%%%%%%%%%%%%%%%%%%%%%%%%%%%%%%%%%%%%%%%%%%%%%%%%%%%%%%%%%%
%
%
\section{Public Blockchain Redactability Limitations}\label{Sec:Limitations}
%
%
%%%%%%%%%%%%%%%%%%%%%%%%%%%%%%%%%%%%%%%%%%%%%%%%%%%%%%%%%%%%%%%%%

Despite their promises, existing blockchain redaction techniques face a number \linebreak of critical
limitations that hinder their practical deployment and long-term \linebreak viability in public blockchains.

One of the foremost concerns is related to security vulnerabilities and trust assumptions. CH-based
approaches often suffer from the aforementioned key exposure problem~\cite{A_Survey_on_Redactable}: if
sensitive information is leaked, the trapdoor key can be recovered, enabling unauthorized redactions. The
management of these keys becomes a delicate issue. Centralized control creates a single point of failure and
compromises decentralization, while distributed key sharing schemes demand significant computational
resources and often lack robust accountability mechanisms\footnote{Effective accountability requires the
ability to identify malicious participants, attribute unauthorized actions to specific parties, and enforce
meaningful penalties. Many distributed CH schemes fail to provide these capabilities, making it difficult to
detect, prove, and punish misuse of redaction privileges.}. Additionally, some decentralized CH schemes have
demonstrated insufficient collision resistance, allowing malicious nodes to manipulate data retroactively
under certain conditions~\cite{Redactable_Blockchain_From_Decentralized}. The absence of effective version
control can lead to reversion attacks, where older, invalid transactions are reintroduced to overwrite
legitimate data. In decentralized environments, there is also the risk of collusion among trapdoor key
shareholders, potentially enabling coordinated misuse of redaction authority without detection or
attribution. Compounding these issues, many schemes fail to preserve prior block states, limiting
transparency and making public auditing impossible.

From a performance perspective, redaction operations can be computationally expensive. Collision finding and
hash recalculations in CH-based systems require significant processing power, with cryptographic techniques
such as multi-party computation or secret sharing only adding to the burden~\cite{A_Survey_on_Redactable}.
Certain designs that involve rewriting entire blocks can cause heavy communication overhead, as nodes must
exchange and validate updated versions of the blockchain. Voting-based redaction mechanisms, particularly in
permissionless settings, tend to suffer from extended confirmation times -- sometimes requiring hundreds of
blocks or multiple days to reach consensus. Storage is another concern. Approaches that append data instead
of deleting it lead to redundant information and wasted space.  Since old data often remains accessible,
even if hidden, scalability is negatively impacted, especially in blockchains that lack pruning mechanisms.

Data consistency and integrity also present significant challenges. In schemes where older data is
overwritten without preserving its previous state -- such as some polynomial or RSA-based models --
historical audits become impossible~\cite{RSA_and_redactable_blockchains}. Moreover, many solutions are
better suited for stateless content and struggle with modifying stateful data like transactions that affect
subsequent outputs or UTXO sets. Poor handling of these dependencies can result in inconsistencies and
unintended cascade effects. In some CH-based models, block hashes do not change after redaction, which may
lead to desynchronization between nodes if updates are not uniformly propagated.

Compatibility and universality further complicate adoption. Many techniques require significant alterations
to the blockchain's core structure or consensus mechanism, making them incompatible with widely used
platforms such as Bitcoin and Ethereum. Retrofitting these blockchains to accommodate redaction
functionality involves high costs, substantial time investment, and computational overhead. Moreover, most
existing proposals are tailored for specific systems or protocols, limiting their applicability across the
diverse landscape of blockchain platforms.

Finally, operational and legal compliance issues remain largely unresolved. The processes for assigning and
enforcing redaction privileges are still ambiguous, posing risks of overreach or unauthorized changes. Even
when data appears to be redacted, remnants may persist in outdated copies of the blockchain, raising doubts
about genuine compliance with data protection laws such as the GDPR. Perhaps most importantly, many of these
technologies are still at the research or prototype stage, with few real-world implementations to validate
their effectiveness or security in practical environments.

%%%%%%%%%%%%%%%%%%%%%%%%%%%%%%%%%%%%%%%%%%%%%%%%%%%%%%%%%%%%%%%%%
%
%
\section{Redactability in Private Blockchains}
\label{Sec:Private_Blockchains}
%
%
%%%%%%%%%%%%%%%%%%%%%%%%%%%%%%%%%%%%%%%%%%%%%%%%%%%%%%%%%%%%%%%%%

Private blockchains are distributed ledgers in which only selected, authorized entities can access data and
participate in the consensus process. This contrasts with public or permissionless blockchains (e.g.,
Bitcoin, Ethereum), where anyone can join, verify, and append data. Permissioned blockchains are typically
governed by a central entity or a consortium, providing enhanced control over data flow, user access, and
system updates.

This control comes at the cost of reduced decentralization but offers considerable advantages in scenarios
where privacy, regulatory compliance, and institutional trust are paramount. For these reasons, private
blockchains are increasingly deployed in enterprise, governmental, and industrial applications. Examples
include:

\begin{itemize}

  \item \textbf{Financial Systems:} Used for interbank transactions, central bank digital currencies
(CBDCs), and regulatory compliance.

  \item \textbf{Supply Chain Management:} Enabling traceability, auditability, and fraud prevention across
trusted participants.

  \item \textbf{Healthcare:} Managing sensitive patient data while ensuring privacy and secure sharing
between trusted institutions.

  \item \textbf{Industrial IoT and Smart Manufacturing:} Collecting, verifying, and occasionally correcting
sensor data within a controlled ecosystem.

  \item \textbf{Identity Management:} Supporting user-centric control over personal data and credentials.

\end{itemize}

In particular, the growing interest from central banks in issuing CBDCs reflects a broader shift toward
permissioned or private architectures. These systems offer scalability and low energy consumption, along
with tools for enforcing rules such as anti money laundering (AML).

Private blockchains are expected to become more prominent in the future. Their ability to balance
decentralization with accountability makes them suitable for next-generation applications in digital
finance, public infrastructure, and digital identity ecosystems. A key component of this evolution is the
redactability property.

The governance of redactable features in private blockchains varies depending on the system’s design. Three
main models are currently explored:

\begin{itemize}

  \item \textbf{Central Authority:} In systems governed by a single entity -- such as a corporation,
regulatory body, or public institution -- the trapdoor key used for enabling redactions is held by that
authority. This model simplifies decision making and execution, but introduces a single point of failure and
potential misuse of power. It is suitable for environments with strong legal oversight or where a single
party bears responsibility (e.g., central banks, healthcare regulators).

  \item \textbf{Consortium-Based Governance:} Here, a pre-approved set of entities jointly manage the
blockchain and the redaction mechanism. The redaction privilege is shared by using cryptographic techniques
like secret sharing or multi-party computation. This model improves resilience and trust, as no single
participant can unilaterally modify the data. It is particularly useful in collaborative environments such
as supply chains, interbank networks, and consortium-led identity platforms.

  \item \textbf{Public Trapdoor:} A more radical approach involves making the trapdoor key public and
embedding it into the blockchain itself (as in the PRBFPT framework proposed
in~\cite{PRBFPT_A_Practical_Redactable}). Rather than relying on designated authorities, all nodes can
verify and potentially initiate redactions, provided that they follow a voting or consensus-based mechanism.
This model aligns with decentralization ideals while supporting redactability, but must be carefully
designed to prevent abuse and maintain system integrity.

\end{itemize}

Regardless of whether redactability is governed by a consortium or through public trapdoor mechanisms, a
supervisory layer involving a designated central authority can be integrated to enhance accountability and
resolve disputes. In this hybrid model, the central authority does not possess unilateral redaction power
but acts instead as an oversight body. It may intervene in exceptional cases, such as contested redactions,
failure of the voting process, or suspected collusion among redaction participants. This approach combines
the benefits of decentralized governance with institutional trust and legal enforceability, thus making it
suitable for regulated sectors.

%%%%%%%%%%%%%%%%%%%%%%%%%%%%%%%%%%%%%%%%%%%%%%%%%%%%%%%%%%%%%%%%%
%
%
\section{Applications and Use Cases}
\label{Sec:Applications}
%
%
%%%%%%%%%%%%%%%%%%%%%%%%%%%%%%%%%%%%%%%%%%%%%%%%%%%%%%%%%%%%%%%%%

Blockchain redactability introduces controlled mutability into traditionally immutable distributed ledgers,
enabling systems to update, remove, or correct previously recorded data. This capability is particularly
critical in application domains where data privacy, regulatory compliance, system scalability, or mutable
state management are essential. Use cases for redactable blockchains span a wide range of sectors, including
finance, healthcare, identity management, industrial IoT, and autonomous systems. In these contexts,
redactable ledgers offer the ability to meet requirements such as the GDPR's RTBF, reduce storage burden
from obsolete data, or correct erroneous entries in smart contracts -- all while preserving overall ledger
integrity and auditability.

Two particularly promising areas for applying redactable blockchains are the Internet of drones (IoD) and
federated learning (FL). We discuss them in the rest of this section (Sections~\ref{SubSec:IoD}
and~\ref{SubSec:FL} respectively), which concludes with a summary of broader applications
(Section~\ref{SubSec:Broader_Applications}).

%%%%%%%%%%%%%%%%%%%%%%%%%%%%%%%%%%%%%%%%%%%%%%%%%%%%%%%%%%%%%%%%%
%
\subsection{Redactable Blockchains in the Internet of Drones}
\label{SubSec:IoD}
%
%%%%%%%%%%%%%%%%%%%%%%%%%%%%%%%%%%%%%%%%%%%%%%%%%%%%%%%%%%%%%%%%%

IoD refers to an emerging distributed architecture designed to manage fleets of drones operating in
coordinated, often autonomous manners across a shared airspace. Analogous to IoT, IoD systems enable aerial
vehicles to interact with ground stations and cloud services to perform tasks such as delivery,
surveillance, environmental monitoring, and smart transportation.

As the number of drones increases -- especially in scenarios like air taxis, autonomous delivery, or
disaster response -- IoD networks face mounting challenges in authentication, secure communication, data
classification, and scalable storage. Security and privacy are particularly pressing, as drones frequently
transmit sensitive, application-specific data across different operational zones. Furthermore, while
offering decentralization and data integrity, traditional blockchain-based authentication systems become
increasingly inefficient due to the immutable and append-only nature of conventional ledgers.

To address these challenges, ReBAS (Redactable Blockchain-Assisted Application-Aware Authentication System)
has been proposed in~\cite{A_Redactable_Blockchain_Assisted}. It is a security and authentication framework
tailored for IoD environments, which combines lightweight cryptographic primitives with redactable
blockchain mechanisms to support secure, scalable, and flexible interactions between drones and ground
stations.

One of the distinctive aspects of ReBAS is its application-aware authentication mechanism. Instead of
relying on a single cryptographic session key for all data exchanges, ReBAS establishes data-type-specific
secret session keys between each drone and the ground station. This approach ensures that sensitive
information from one application -- such as surveillance -- is not exposed during the execution of another
-- like package delivery.

To support the computational constraints of drones, ReBAS adopts Chebyshev polynomials as the basis for its
lightweight cryptographic operations. These mathematical functions facilitate efficient key generation and
mutual authentication without burdening the limited processing resources of aerial devices.

ReBAS also leverages a consortium blockchain architecture, implemented via Hyperledger
Fabric~\cite{Hyperledger_Fabric}, where ground stations collaboratively maintain a distributed ledger. This
ledger stores drone identities, cryptographic credentials, and assigned missions, in addition to being
specifically designed to support controlled modifications. The inclusion of CH functions allows authorized
parties to modify specific ledger entries -- such as when reassigning a drone's task -- without invalidating
the blockchain’s structure or compromising its integrity.

This integrated design enables ReBAS to meet the evolving needs of IoD systems by supporting dynamic
reconfiguration, secure data handling, and long-term scalability, all while maintaining the verifiability
and trustworthiness inherent to blockchain technology.

The redactable blockchain design provides two critical functionalities:

\begin{itemize}

  \item \textbf{Efficient Task Reassignment:} When a drone is reassigned to new tasks (e.g., changing from
traffic monitoring to delivery), its ledger record can be securely updated by using CH collisions.
    
  \item \textbf{Storage Optimization:} By allowing updates rather than continuous appending, the system
avoids exponential growth in stored data, addressing long-term storage and scalability concerns.

\end{itemize}

The integration of redactable blockchains within ReBAS proves to be instrumental for modern IoD systems.
First, redactability supports frequent and dynamic updates to drone assignments and data policies without
bloating the ledger. Second, it ensures compliance with data protection regulations by allowing sensitive or
erroneous data to be altered or erased. Third, redactable ledgers empower cross-zone coordination among
ground stations without incurring prohibitive overhead or compromising system integrity.

%%%%%%%%%%%%%%%%%%%%%%%%%%%%%%%%%%%%%%%%%%%%%%%%%%%%%%%%%%%%%%%%%
%
\subsection{Redactable Blockchains in Federated Learning}
\label{SubSec:FL}
%
%%%%%%%%%%%%%%%%%%%%%%%%%%%%%%%%%%%%%%%%%%%%%%%%%%%%%%%%%%%%%%%%%

FL is a distributed machine learning paradigm in which multiple edge devices or clients collaboratively
train a shared model while keeping their local data private. This approach is particularly advantageous in
environments such as the industrial Internet of things (IIoT), where industrial sensors and machines
generate proprietary or sensitive data. FL allows these entities to contribute to a global model without
transmitting raw data, thereby preserving privacy and reducing communication overhead.

Despite its advantages, FL presents several challenges, including how to verify model updates, ensure data
provenance, protect against poisoning attacks, and maintain auditability across decentralized stakeholders.
While traditional blockchain systems have been proposed to address these concerns by offering an immutable
record of model updates, their rigidity introduces limitations. For instance, invalid or malicious
contributions cannot be removed once recorded; moreover, the growing ledger size can hinder scalability.
Redactable blockchains offer a compelling alternative by preserving data integrity and traceability while
supporting controlled modifications.

To address the limitations of immutable logging in FL systems, in~\cite{A_Redactable_Blockchain_Framework} a
framework has been proposed that integrates redactable blockchain mechanisms into the FL pipeline within
IIoT contexts. This hybrid system maintains the core benefits of decentralized auditing and verifiability,
while adding flexibility for data correction and deletion.

The framework employs a permissioned blockchain to log all local model updates submitted by participating
IIoT devices. Each device trains locally and submits encrypted updates to the ledger. This ensures
decentralized accountability and verifiable model provenance.

CH functions are embedded into the blockchain to enable selective redaction of ledger entries. Authorized
parties can perform controlled hash collisions to replace or remove previously submitted updates without
breaking the chain structure. This is especially useful for eliminating faulty or adversarial model
contributions.

In this framework the FL process unfolds over three coordinated phases:

\begin{enumerate}

  \item \textbf{Registration and Authentication:} IIoT clients and devices are authenticated and registered
on the blockchain network.

  \item \textbf{Model Update Logging:} After local training, clients encrypt and submit their updates to the
ledger.

  \item \textbf{Aggregation and Validation:} Edge servers validate the received updates, aggregate them, and
distribute global model parameters.

\end{enumerate}

If an edge server detects anomalies -- such as model poisoning or outdated updates -- it can trigger a
redaction request by using its trapdoor key. The system updates or invalidates the affected ledger entry
without requiring chain reorganization or rollback, thereby preserving system continuity.

Experimental evaluations confirm the framework's ability to maintain high model accuracy and convergence
while reducing the storage overhead typical of immutable blockchain solutions. The design remains scalable
and energy-efficient, which is essential in resource-constrained IIoT settings.

Redactability introduces critical enhancements to FL systems, particularly in industrial scenarios where
resilience, accuracy, and regulatory compliance are paramount. Unlike traditional blockchains, where
poisoned or faulty updates become permanently embedded in the ledger, redactable blockchain systems offer
the flexibility to revise or remove problematic contributions post hoc. This capability ensures that model
integrity is preserved even in the presence of adversarial or erroneous updates.

Moreover, redactability supports privacy-conscious debugging by allowing the redaction of sensitive
information inadvertently exposed through model contributions, thereby aligning system behavior with data
protection regulations. It also enables more efficient storage management: outdated or redundant updates can
be pruned from the ledger, reducing data accumulation and improving scalability over time.

In conclusion, redactable blockchains provide the dynamic adaptability required to deploy FL at industrial
scale. By supporting secure, auditable, and flexible update management, they ensure that FL systems can
uphold performance, maintain data integrity, and meet compliance obligations in real time, making them
viable for complex IIoT applications.

%%%%%%%%%%%%%%%%%%%%%%%%%%%%%%%%%%%%%%%%%%%%%%%%%%%%%%%%%%%%%%%%%
%
\subsection{Summary of Broader Applications}
\label{SubSec:Broader_Applications}
%
%%%%%%%%%%%%%%%%%%%%%%%%%%%%%%%%%%%%%%%%%%%%%%%%%%%%%%%%%%%%%%%%%

Beside IoD and FL, redactable private blockchains can be applied in domains where both data transparency and
controlled mutability are necessary:

\begin{itemize}

  \item \textbf{Healthcare:} Managing sensitive patient data, enabling error rectification, and ensuring
compliance with privacy regulations such as HIPAA\footnote{The Health Insurance Portability and
Accountability Act (HIPAA) is a US federal law that establishes standards for protecting sensitive patient
health information, requiring secure handling of protected health information (PHI) and granting patients
rights to access and request corrections to their medical records.} and GDPR.

  \item \textbf{Digital Identity:} Supporting revocation, updates, and secure management of credentials
while preventing unauthorized or fraudulent access.

  \item \textbf{Finance:} Powering features like reversible tokens to address theft or error, auditing
regulatory reporting, and facilitating consolidation of financial records.

\end{itemize}

Across all these sectors, the general need for controlled data mutability in permissioned (private or
consortium) settings is clear. Redactable blockchains enable compliance with evolving regulations,
rectification of errors or malicious entries, and adaptability to real-world operational demands -- making
them an essential evolution in secure, trustworthy, and practical blockchain-based systems.

%%%%%%%%%%%%%%%%%%%%%%%%%%%%%%%%%%%%%%%%%%%%%%%%%%%%%%%%%%%%%%%%%
%
%
\section{Conclusion}
\label{Sec:Conclusion}
%
%
%%%%%%%%%%%%%%%%%%%%%%%%%%%%%%%%%%%%%%%%%%%%%%%%%%%%%%%%%%%%%%%%%

Redactable blockchains represent a significant advancement in distributed ledger technology, as they address
the practical limitations of strict immutability. By enabling controlled and auditable modifications, these
systems reconcile the foundational benefits of blockchain -- transparency, security, and trustworthiness --
with emerging regulatory, operational, and scalability requirements. Central to this evolution are
cryptographic primitives such as chameleon hash functions, alongside alternative approaches like verifiable
delay functions and zero-knowledge proofs. 

Redactable blockchains are proving especially valuable in permissioned environments such as finance,
healthcare, IIoT, and emerging domains like FL and IoD. Their ability to balance ledger integrity with
controlled mutability paves the way for broader blockchain adoption in compliance-sensitive and dynamic data
ecosystems.

Continued research and development are needed to mature these techniques, improve their trust models, and
expand real-world deployment. Nevertheless, redactable blockchains offer a pragmatic and forward-looking
solution for building trustworthy digital infrastructures responsive to both technological and societal
demands.

%ZZZ (forward consequence propagation problem)
Despite ongoing efforts related to secure key management, performance, and system compatibility, several
fundamental problems remain open. A particularly critical issue is the challenge of tracking and managing
the forward propagation of consequences from redacted transactions: \emph{what are the transactions
depending on redacted ones and how should they be redacted in turn?} Unlike reversible
computing~\cite{Lan61,Ben73,BAPCDL12,Fra18}, where \emph{backward} procedures can systematically undo
computations~\cite{DK04,Kri12,LMS10,CKV13,PU07a,LMM21,LPU24,BM23a,BLMY24,BM23b,BM24,DMV90,Bed91,BR23,BEM24},
blockchain redactions face the complex task of identifying and appropriately handling all dependent
transactions and state changes that resulted from the original, now-modified data. This \emph{forward}
consequence management represents a fundamentally different computational paradigm that requires novel
approaches to maintain system consistency and integrity.

\medskip
\noindent
\textbf{Acknowledgments.} This research has been supported by the PRIN 2020 project \emph{NiRvAna --
Noninterference and Reversibility Analysis in Private Blockchains}. The scholarship of the first author at
the Italian PhD Program in Blockchain and Distributed Ledger Technology has been funded by \emph{PNRR --
Piano Nazionale di Ripresa e Resilienza} according to D.M.~118/2023.

\bibliographystyle{unsrt}
\bibliography{biblio}

\end{document}